\documentclass[twocolumn,showpacs,aps,prl]{revtex4}  
\usepackage{epsfig,amssymb}

\begin{document}

\title{Stationary and non-stationary fluid flow of a Bose-Einstein condensate through a penetrable barrier}

\author{P. \surname{Engels}}
\email{engels@wsu.edu}
\author{C. \surname{Atherton}}
\affiliation{Washington State University, Department of Physics
and Astronomy, Pullman, Washington 99164, USA}

\date{\today}

\begin{abstract}
We experimentally study the fluid flow induced by a broad,
penetrable barrier moving through an elongated dilute gaseous
Bose-Einstein condensate. The barrier is created by a laser beam
swept through the condensate, and the resulting dipole potential
can be either attractive or repulsive. We examine both cases and
find regimes of stable and unstable fluid flow: At slow speeds of
the barrier, the fluid flow is stationary due to the superfluidity
of the condensate. At intermediate speeds, we observe a
non-stationary regime in which the condensate gets filled with
dark solitons. At faster speeds, soliton formation completely
ceases and a remarkable absence of excitation in the condensate is
seen again.
\end{abstract}

 \maketitle

\par
The fluid flow past an obstacle is one of the most prototypical
experiments studying superfluidity. Accordingly, moving an
obstacle through a superfluid has been met with considerable
interest, both in the context of superfluid helium
\cite{Donnelly1991} as well as in the context of dilute gaseous
Bose-Einstein condensates (BECs) \cite{Leggett2001}. In
experiments with BECs, the role of the obstacle can be played by a
laser beam that creates a dipole potential for the atoms. By
moving a small, strongly repulsive dipole beam through a BEC,
evidence for a critical velocity above which superfluidity breaks
down was obtained \cite{Raman1999,Onofrio2000}. In those
experiments, the laser was impenetrable for the atoms, and the
diameter of the laser beam was chosen smaller than the size of the
BEC. The atoms could therefore flow past the sides of the moving
beam. The critical velocity observed in the experiments was much
lower than that given by the well-known Landau criterion
\cite{Landau1941}. Vortices can be produced by such narrow
obstacles, and indeed they were observed experimentally
\cite{Inouye2001}, and in numerical simulations
\cite{vortexnumerics}. Likewise, in the context of optical
lattices, heating due to dissipative motion has been used as a
tool to study superfluidity in the presence of periodic potentials
\cite{lattice}.

\begin{figure} \leavevmode \epsfxsize=3.375in
\epsffile{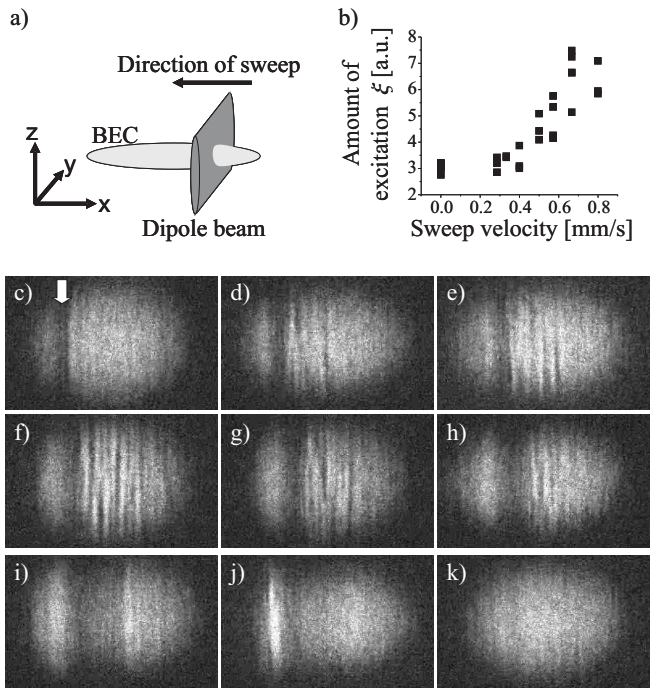} \caption{\label{sweepspeed} (a) Axis
convention. The dipole beam propagates in the -y direction and is
moved through the BEC along the -x direction. (b) Onset of
excitation at slow sweep speeds, calculated from images with
parameters as in (c)-(k). A repulsive dipole beam is swept through
the cloud from the right to the left. The white arrow in Fig. (c)
indicates the end position. Sweep speeds are (c)~0.4~mm/s,
(d)~0.5~mm/s, (e)~0.6~mm/s, (f)~0.7~mm/s, (g)~0.8~mm/s,
(h)~1~mm/s, (i)~1.3~mm/s, (j)~2~mm/s, (k)~3.3~mm/s. Images show a
200~$\mu m$ wide region. Other parameters see text.}
\end{figure}

In this Letter we consider a situation that is complementary to
the experiments described in \cite{Raman1999,Onofrio2000}: In our
case, the BEC is contained in a narrow, elongated trap and a
barrier much wider than the radial extent of the BEC is moved
through it (Fig.~\ref{sweepspeed}(a)). The atoms cannot flow past
the sides of the barrier, and we use a penetrable barrier allowing
atoms to flow \textit{through} the barrier. The barrier is created
by the dipole potential of an elliptically shaped laser beam that
is swept along the long axis of the elongated BEC. Such a geometry
approximates a one-dimensional problem, and it is theoretically
expected that in this case solitons play a similar role to the
role played by vortices in higher dimensions
\cite{Hakim1996,Pavloff2002,Radouani2004,Astrakharchik2004}.
Recently, shedding of solitons has also been discussed in terms of
Cerenkov-radiation \cite{El2006,Carusotto2006,Susanto2007}. In
this Letter we provide experimental evidence for the existence of
several different flow regimes in the wake of the moving barrier.
For slow and for fast velocities, the wake behind the moving
barrier appears unperturbed. In an intermediate velocity regime
solitons appear. This behavior is observed with repulsive barriers
as well as with attractive potentials.

The starting point for the experiments described in this paper are
elongated $^{87}$Rb BECs of about $4.5\cdot10^{5}$ atoms in the
$|F=1,m_{F}=-1\rangle$ state, held in a Ioffe-Pritchard type trap
with trapping frequencies
$\{\omega_{x}/(2\pi),\omega_{yz}/(2\pi)\}=\{7.1,203\}$Hz. The axis
of weak confinement (x-axis, see Fig.~\ref{sweepspeed}(a)) is
oriented horizontally. Evaporative cooling is performed until no
thermal cloud of uncondensed atoms is visible. A laser that is far
detuned from the Rb absorption lines is directed along the second
horizontal axis (y-axis). Depending on the detuning, it can create
an attractive or repulsive dipole potential for the atoms. This
dipole beam can be moved along the x-axis at various speeds,
inducing flow in the BEC. The dipole beam waist in the vertical
z-direction is much larger than the radial extent of the BEC so
that the intensity variation of the beam along the z-direction can
be neglected. The imaging direction coincides with the direction
of the dipole beam. A notch filter prevents the dipole beam from
being detected by the camera.

As we show in this Letter, solitons can be produced by sweeping
the dipole beam through the condensate. The width of these
solitons is below our spatial imaging resolution when the BEC is
held in the trap. Therefore we employ a fast antitrapping
technique to expand the BEC before imaging, similar to the
procedure described in \cite{Lewandowski2003}. The expansion
imaging procedure starts with a microwave adiabatic rapid passage,
converting the atoms from the $|F=1,m_{F}=-1\rangle$ state to the
antitrapped $|F=2,m_{F}=-2\rangle$ state. We then set the magnetic
bias field to 48~G and let the cloud  expand for 3~ms. During this
expansion, the magnetic field is positioned such that the
antitrapping potential keeps the cloud from falling under the
influence of gravity. The atoms are then imaged in a 100~G bias
field using resonant absorption imaging. The antitrapping
predominantly leads to a radial expansion, so that the aspect
ratio of the BEC changes from $R_{yz}/R_{x}$ = 29 for the trapped
BEC to a value of about 1.6 at the end of the expansion time.

In a first set of experiments, shown in Fig.~\ref{sweepspeed}, we
study the flow induced by a weak repulsive potential swept through
the BEC at various speeds. The potential was produced by a laser
beam with a wavelength of 660~nm and a power of 37~$\mu W$. The
beam shape was elliptical with waists $w_{x} = 7.6~\mu m$ and
$w_{z} = 25.8~\mu m$. Such a laser produces a weak repulsive
potential whose height is about 24\% of the chemical potential of
the condensate. The healing length in the center of our typical
condensates is 0.17~$\mu m$ so that the dipole beam can be
considered wide with respect to the healing length even in the
direction of the sweep. The dipole beam starts out on the right,
outside the BEC, and is then swept towards the left at a constant
speed as given in the figure caption. The sweep is stopped at the
position indicated by the white arrow in Fig.~\ref{sweepspeed}(c),
and the BEC is imaged after antitrapped expansion as described
above.

We can identify several different regimes, depending on the sweep
speed. For very slow sweep speeds, such as in
Fig.~\ref{sweepspeed}(c), the part of the BEC through which the
dipole beam has been swept appears essentially unaffected, and the
only visible effect of the beam is the density suppression that it
leaves at its end position. Since the BEC is superfluid, it is not
surprising that it remains unaffected by a slowly moving
penetrable barrier. At faster sweep speeds
(Fig.~\ref{sweepspeed}(d)-(g)), dark notches appear in the wake of
the dipole beam. We interpret these notches as the solitons
predicted in theoretical studies of the one-dimensional problem
\cite{Hakim1996,Pavloff2002,Radouani2004,Astrakharchik2004,Susanto2007}
(see below for an experimental verification). Some of the notches
observed in the wake of the dipole beam show slight bending or
nonuniform contrast along the plane, while others are very
straight and uniform. In the darkest solitons, we observe a
suppression of the central density by about 50\%. However, it is
conceivable that the true depth of some of the solitons is larger
than this because slight misalignments of the imaging direction
with respect to the soliton planes can reduce the apparent depth.
Interestingly, fewer solitons are observed when the sweep speed is
increased beyond 0.9~mm/s (Fig.~\ref{sweepspeed}(h). This effect
is accompanied by the appearance of an increasingly wider region
of low density directly behind the dipole beam
(Fig.~\ref{sweepspeed}(h),(i)). At faster sweep speeds, the
condensate appears again unaffected by the obstacle, as seen in
Fig.~\ref{sweepspeed}(k). The apparent absence of excitations in
the wake of the beam is very striking. Suppression of phonons and
solitons at high obstacle velocities was theoretically discussed
in \cite{Law2000,Pavloff2002,Radouani2004}. In particular,
\cite{Radouani2004} describes an explanation of related numerical
findings by using an analogy to the radiation of capillary-gravity
waves in a classical fluid.

To quantify the amount of excitation in the BEC, we first
calculate the smooth, overall shape of each BEC by considering
only the lowest Fourier components of each image. Then the
root-mean-square deviation of each original image from the smooth,
overall shape is used as a measure of the excitation present in
the BEC. To reduce the influence of the inhomogeneous density
distribution along the x-axis, we only consider a region in the
center of the BEC that is 35~$\mu m$ wide in the x-direction and
70~$\mu m$ wide along the z-direction. The resulting quantity,
$\xi$, is plotted versus the barrier speed in
Fig.~\ref{sweepspeed}(b). The points at zero velocity are taken
from unperturbed BECs in the absence of a dipole beam, and mostly
indicate imaging noise. The onset of excitations above a critical
velocity of roughly 0.3~mm/s can clearly be seen in this plot. We
do not extend this plot to velocities above 0.8~mm/s because at
higher speeds the broad low-density region that develops in the
wake of the barrier (as seen in Fig.~\ref{sweepspeed}(h), (i))
affects our measure. For comparison, the bulk speed of sound in
the center of our BECs is 3~mm/s. In a cigar shaped BEC, averaging
over the tightly confined direction leads to a lower speed of
sound \cite{Zaremba1998,Stringari1998,Kavoulakis1998}. In our
case, this results in a speed of 2.1~mm/s which is still
significantly larger than the observed critical velocity. This
situation is reminiscent of the observation of a surprisingly low
critical velocity in vortex shedding experiments in
\cite{Raman1999,Onofrio2000}. For a more precise analysis, the
complicated fluid flow through the barrier would have to be taken
into account. In addition, the nonuniform radial density profile
of the BEC is expected to lead to a deviation from the Bogoliubov
dispersion relation and to a lower critical velocity, as discussed
in \cite{Fedichev2001}.

\begin{figure} \leavevmode \epsfxsize=3.375in
\epsffile{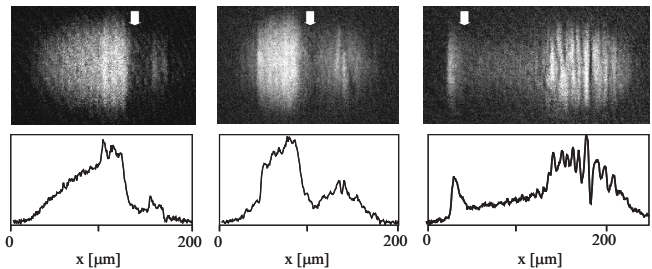} \caption{\label{xsections} Images (upper
row) and corresponding integrated cross sections (lower row) for
various endpoints of the sweep as indicated by the white arrows.
Laserpower 69~$\mu W$, sweep velocity 1.4~mm/s.}
\end{figure}

When a more strongly repulsive potential is used, we observe
similar behaviors. However, solitons appear already at lower sweep
velocities, and the velocity above which the flow becomes
stationary again increases. Also, a more pronounced accumulation
of atoms in front of the dipole beam is seen, delimited by a steep
wavefront (Fig.~\ref{xsections}). For weaker potentials the number
of solitons in the BEC is reduced. For example, when using dipole
beams with an intensity of 25~$\mu W$, often only one or two
individual solitons are produced.

\begin{figure} \leavevmode \epsfxsize=3.375in
\epsffile{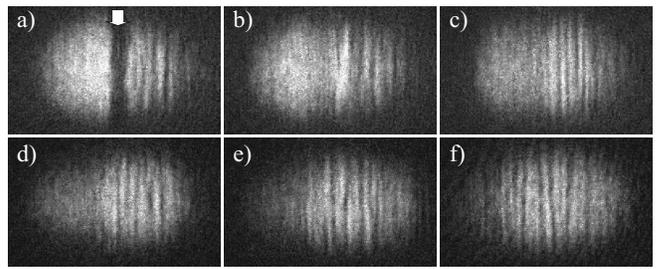} \caption{\label{evotime} Stability of the
observed solitons. Times for which the BEC is allowed to evolve in
trap after turning the dipole beam off at the end of the sweep and
before starting the expansion imaging procedure: (a)~0~ms, (b)~
10~ms, (c)~20~ms, (d)~30~ms, (e)~40~ms, (f)~50~ms. The arrow in
(a) indicates the end position of the sweep. Other parameters see
text. Images show a 200~$\mu m$ wide region.}
\end{figure}

In a second set of experiments we now show that the dark notches
in the wake of the dipole beam are not just soundwaves but do
indeed behave like solitons. To prove this, we compare the
stability of the observed notches with the stability of a simple
density suppression in the BEC (Fig.~\ref{evotime}). To obtain a
very clearly visible density suppression, a relatively strong
dipole beam with a power of 129~$\mu W$ was swept through half of
the BEC. The height of the dipole potential was about 85\% of the
chemical potential of the condensate so that a deep density
suppression was left in the BEC at the end position of the sweep.
 For clarity this suppression is marked by the white arrow in Fig.~\ref{evotime}(a).
The sweep velocity was adjusted to 0.2~mm/s, leading to pronounced
soliton formation in the wake of the laser. At the end of the
sweep, the dipole beam was switched off, and the BEC was allowed
to evolve in trap for a variable evolution time before starting
the expansion imaging procedure. Typical images for various
evolution times are shown in Fig.~\ref{evotime}. At very short
evolution times (Fig.~\ref{evotime}(a)), the right part of the
cloud, through which the dipole beam has been swept, shows stacks
of dark notches, and the density suppression produced by the
dipole beam at its end position is clearly visible. This density
suppression gets filled in by the BEC over a time scale of about
10~ms (Fig.~\ref{evotime}(b)). Considering that the longitudinal
speed of sound in our cigar-shaped BEC is 2.1~mm/s, it is very
plausible that the density suppression with an initial width of
roughly 20~$\mu m$ closes during 10~ms. However, the much narrower
dark notches in the right part of the cloud remain visible for
much longer times and slowly spread out
(Fig.~\ref{evotime}(b)-(f)). After evolution times of over 50~ms
they are still visible and have spread over the entire BEC
(Fig.~\ref{evotime}(f)). We often observe very straight dark
notches at this time. Then their contrast gradually fades and they
disappear. These measurements demonstrate the stability of the
produced dark notches, suggesting that they are indeed very
distinct from sound waves and correspond to the solitons predicted
by theory for the one-dimensional problem.

Our condensates are elongated, but not one-dimensional. Therefore,
the question arises which influence the transverse dimension has
on the existence of solitons. In previous experiments, individual
solitons in a BEC with repulsive interactions have been engineered
using phase imprinting \cite{Burger1999,Denschlag2000} or
wavefunction engineering \cite{Anderson2001}. They have also been
observed in the evolution of a density gap in a BEC produced by a
stopped light technique \cite{Dutton2001}. One important outcome
of the phase imprinting studies is that the exact nature of the
solitary texture depends on the strength of the radial trapping.
The numerical studies given in \cite{Muryshev2002} reveal that if
the chemical potential exceeds ten times the level spacing in the
tightly confined direction, i.e. if $\mu /\hbar \omega_{yz}> 10$,
phase imprinting produces a nonstationary kink that decays into
phonons on a time scale of $1/\omega_{yz}$. If this ratio is below
five, the kink decays into a proper dark soliton. Experimentally,
solitons produced by phase imprinting were found in experiments
with $\mu /\hbar \omega_{yz} = 7$ \cite{Burger1999}. While our
mechanism for soliton formation is not based on phase imprinting,
a similar influence of the transverse confinement on the existence
of solitons can be expected. In particular, for our BECs $\mu
/\hbar \omega_{yz} = 9.7$, which is not too different from the
experiments in \cite{Burger1999}.

\begin{figure} \leavevmode \epsfxsize=3.375in
\epsffile{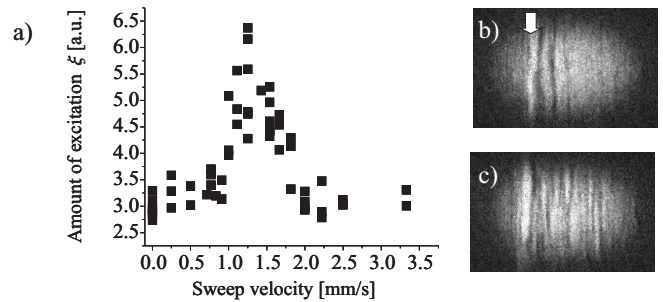} \caption{\label{excitation830} Effect of
an attractive potential swept through the condensate. (a)
Excitation strength versus sweep speed for a laser power of
10~$\mu W$, other parameters see text. (b)~Laser power 10~$\mu W$,
sweep speed 1.25~mm/s. (c)~Laser power 19~$\mu W$, sweep speed
1.25~mm/s. Images show a 200~$\mu m$ wide region. The white arrow
in (b) indicates the end position of the sweep.}
\end{figure}

By changing the laser wavelength to 830~nm we can also study the
effects of an attractive potential, in which case the density is
increased instead of decreased in the region of the potential. As
in the case of repulsive potentials, theories predict soliton
formation at intermediate speeds and vanishing excitation at very
slow and very fast speeds \cite{Pavloff2002}. We indeed observe
such regimes. However, deviating from our observations with
repulsive potentials, we have not found a region of suppressed
density behind an attractive dipole laser that would correspond to
Fig.~\ref{sweepspeed}(i). Results obtained with an attractive
dipole beam are shown in Fig.~\ref{excitation830}. A dipole beam
with a wavelength of 830~nm and a waist of 8.2~$\mu m$ in the
x-direction and 31.7~$\mu m$ in the z-direction was used. For the
analysis in Fig.~\ref{excitation830}(a), a laser power of 10~$\mu
W$ was chosen, leading to a potential depth of about 17\% of the
chemical potential of the BEC. As before we quantify the
excitation by plotting $\xi$ versus the sweep speeds. Since we do
not observe a low-density region corresponding to
Fig.~\ref{sweepspeed}(i), we can now use $\xi$ as a measure for
the excitation over the whole range of velocities studied. Despite
the weakness of the perturbing potential, soliton formation again
occurs at surprisingly low velocities.

For a laser power of 10~$\mu W$, only a few solitons are produced,
as shown in Fig.~\ref{excitation830}(b). If the laser power is
increased to 19~$\mu W$, the soliton formation can become quite
pronounced (Fig.~\ref{excitation830}(c). This demonstrates that
not only a repulsive potential, but also an attractive potential
can be an efficient tool to produce solitons. One-dimensional
theories for the case of attractive potentials predict the onset
of soliton formation to occur right at the speed of sound
\cite{Pavloff2002}. However, we see that already at a speed of
1.25~mm/s, below the central speed of sound, the central part of
the BEC can be filled with solitons. This discrepancy may hint at
an influence of the inhomogeneous density across the condensate in
our experiments.

\par
In conclusion, we have observed stationary and non-stationary
fluid flow regimes when a penetrable barrier is moved through a
BEC. A slowly moving barrier leads to stationary flow. At
intermediate velocities, the flow is non-stationary and is marked
by the emergence of solitons in the wake of the barrier. At fast
velocities, the soliton shedding stops and the flow appears
stationary again. In a one-dimensional geometry, the solitons play
a role analogous to vortices present in flow around obstacles in
higher dimensions
\cite{Raman1999,Onofrio2000,Inouye2001,vortexnumerics}. Thus our
experiments complement these previous studies that have been
hallmark experiments for investigating the superfluidity in dilute
gaseous BECs.

\begin{acknowledgments}
\end{acknowledgments}

\end{document}